# Smart packaging of electronics and integrated MEMS devices using LTCC

Suhas Kumar and A. M. Vadiraj

*Abstract*— Low Temperature Cofired Ceramics (LTCC) has been a popular multi layer ceramic (MCM) packaging material for many electronic applications. The main advantage with LTCC would be its ability to embed a major part of the electronic circuit within itself, apart from its enhanced RF functionality as against many lossy materials used. The advantages of LTCC in terms of frequency response, cost, ease of fabrication, etc over many other packaging materials are presented. The applicability of LTCC as a packaging material, circuit mounting material, substrate material or a base material for micro devices is discussed.

Switches and filters fabricated on LTCC as a substrate are presented and their enhanced functionality is shown. Planar switches and RF MEMS switches on LTCC are discussed with regard to their isolation, insertion losses, return losses, repeatability, quality factor, parasitic effects and frequency response. Concern is also shown to parameters like actuation voltages, actuation times and complexity of fabrication. The parameters studied with design and fabrication of filters is also discussed, like Q factor, dispersive effects, limits on frequencies, etc.

Discussion is also done with regard to LTCC as a base material for MEMS sensors and actuators and the performance variables of the same.

Fabrication process parameters are presented. The important issue of feasibility of integration with microelectronic integrated circuitry is discussed and its effects are shown.

## I. INTRODUCTION TO LTCC

LTCC tape systems are composed of a glass/ceramic dielectric tape provided in rolls with shrinkage-matched metallization pastes. Unlike thick film technology where sequential printing, drying, and firing of each layer are required, LTCC technology, like PWB technology, processes all the different layers in parallel (ie. all layers are punched, printed, and dried in parallel). As a result optimum yields and cost effectiveness are achieved since each process is not dependent on the previous process. Each layer can be inspected prior to stacking allowing for extremely high yields. Once all the layers are processed, they are stacked, laminated and co-fired at temperatures between 8OOC-9OOC to form a high density, fully integrated substrate, packaged by adding mechanical leads or seal rings and in some cases thermal heat sinks, BGA (ball grid array) PGA (pin grid array). MPGA are readily accomplished, this can be used at a high cost savings over the more conventional integrated package, separate ceramic or laminate insert that is then mounted in a standard LTCC package which results in additional wire bonding and lower reliability because the I/O's from the substrate to the package are eliminated. During the firing process, the tape will shrink between 12% - 16% + 0.2% in the X and Y-axis and 15%-25%+0.5% in the Z-axis (depending on manufacturer). Assuming balanced metal loading and good process control, the shrinkage is consistent; however, with certain systems, there is now a narrow processing window which the fabrication must take place. [1]

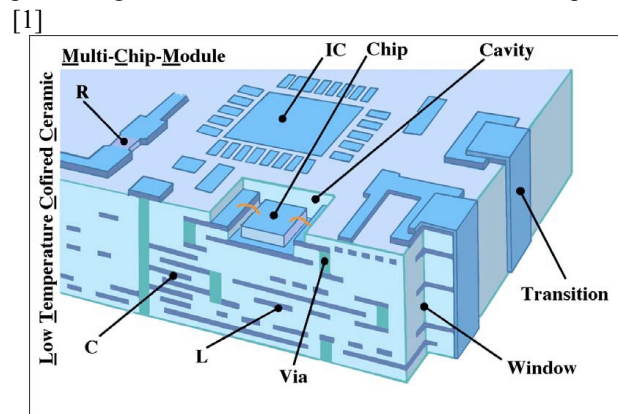

**Figure 1:** A typical LTCC circuit schematic

## II. MCM PACKAGING USING LTCC TECHNOLOGY

Besides cost effectiveness and integrated packaging, LTCC technology offers other features that are ideal for MCM packages.
- Fine line and Spaces
- Low Resistance Metallization
- Excellent High Frequency Characteristics
- Allows for Unique Designs
- Fine vias

### A. Fine Lines and Spaces

While the definition of fine lines and spaces has changed with each improvement of technology, the thick film printing process in LTCC technology is easily capable of 0.004" lines and spaces, and 0.003" lines and spaces are do-able. Finer lines and spaces, using photo image are available to provide line widths and spaces in the order of 0.0015"-0.002". [2]

Suhas K is with the Department of Electronics and Communication Engineering, BMS College of Engineering, Bull Temple Road, Basavanagudi, Bangalore-19, India (e-mail: suhas.asulikeit@gmail.com).

Vadiraj A M is with the Department of Electrical and Electronics Engineering, BMS College of Engineering, Bull Temple Road, Basavanagudi, Bangalore-19, India (e-mail: vadisam@gmail.com).

A thin film deposition process can be used on the outer layers to produce finer lines and spaces: however, it's very expensive, and the surface of the fired tape must be extremely smooth for good adhesion. Polishing will help, but very small pits can remain and cause some adhesion problems. Another method to achieve finer lines and spaces on the top layer is a photo imagable thick film post-fire process. This is less expensive than thin film and two paste manufacturers often this approach enables one to resolve 0.00 1 /0.002" lines and spaces, respectively. The material systems require printing a gold area where the fine resolution is required. After drying, the fine line pattern is exposed onto the gold area by a high intensity ultraviolet light source. The pattern is developed using a sodium carbonate spray process followed by a water rinse and drying. The resulting pattern is then fired through a belt furnace at 850C like a standard thick film print. Other thick film attachable gold processes are available, but the sodium carbonate developer used in this method does not attack the glass in the LTCC, and so no additional masking and stripping is required. A recent development by another paste (thick film) company allows the fine lines and spaces on every conductive layer of the circuit by using a method called TOS (tape on substrate). While fine lines and spaces are available with LTCC technology, it comes at a significant cost. One must weigh the option of adding layers and using wider lines/spaces or using the more expensive thin film or photo printable thick film processes. In many cases where fast speeds or high frequencies are used, there is justification for lines and spaces: however, in the majority of applications, additional layers and wider lines and spaces can be used and significant cost savings can be realized.

### B. Low Resistance Metallization

What makes LTCC technology desirable for many applications is the use of high conductivity metals such as gold and silver. MCM applications requiring high speed digital and/or low level, high frequency analog processing cannot tolerate the IR losses associated with the metallization used in high temperature co-fired ceramic (HTCC) technology. But the metallization in HTCC do offer many cost advantages and in many high impedance applications they have more than adequate conductivity. [3]

Without sacrificing the electrical performance, LTCC tape manufacturers have addressed the cost issue by offering mixed metallization system and a copper system. Mixed metallization systems use lower cost silver in the buried layers and gold is used on the top or bonding layers using a barrier via fill connecting the outer and inner layers, thus eliminating a Kirkendal electrical open. Lower cost copper systems are available with LTCC. While the metallization is lower cost, the fabrication process requires a wet nitrogen process which can be costly to install and maintain if not supported by significant production. The manufacturers of LTCC tape systems will continue to work with their metallization to reduce costs; however, they will always be subject to the cost fluctuations of the world commodity market. As a result, emphasis will be placed on mixed metallization system of gold/silver, copper systems and mix of copper/silver/gold that is solder able and wire bondable. [4]

### C. Excellent High Frequency Characteristics

LTCC tape systems exhibit excellent high frequency characteristics making it ideal for applications (such as Ferro A6). This material has a low dielectric constants, low insertion loss, and good loss tangent characteristics. In addition, the tape is cast in consistent 0.005"/0.008" thickness, so dielectric thickness are predictable to within 5 0.0002". While the low dielectric tape is desired for impedance control, many tape manufacturers are combining the low dielectric with a very high K tape. This enables designers to print high value capacitors on the inner layers of the substrate, thereby as well as resistors and inductors allowing many GaAs MMIC devices to be smaller. Efforts continue in these tape systems for both military and commercial applications.

Since LTCC technology is expected to be used in many GaAs MMIC packaging applications, it is noteworthy that LTCC's CTE closely matches that of GaAs (7.0 vs 6.5) [5]

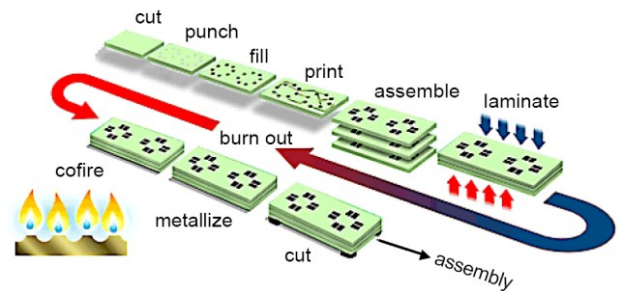

**Figure 2:** A typical LTCC process flow

### III. LTCC FOR MICROSYSTEMS

A characteristic feature of LTCC is good workability. In some cases a LTCC-based microsystem can be a good alternative to microsystems made in silicon or other technologies. Reasons for choosing LTCC-Technology may be financial considerations or specific material properties. A main problem is to simplify a mechanical component in such a way, that it is possible to integrate this component in a planar structure with a small height in consideration of the restrictions of the LTCC Technology. In contrast to LTCC-based substrates with only electrical circuits the integration of mechanical components make other demands on the different technological steps of the LTCC-Process. In this section some 3D-structures made in LTCC-like fluidic channels, membranes usable for micro pumps or pressure sensors – and some aspects of required special technological demands are described. [6-7]

### A. Why LTCC for Microsystems

The LTCC-Technology is a commonly used technology for automotive, RF and other applications with high demands on packaging density and reliability. In nearly all of these applications LTCC-Substrates work as a pure carrier with only electrical functions.

But the special features of the manufacturing process allow it to integrate non-electrical functions very effectively. An important advantage is the possibility of processing LTCC in the "green", non-sintered condition. The design and the structuring of the ceramic is relocated from the ceramic manufacturer to the substrate maker. It allows us to design the ceramic substrate very closely to the application. [8]

Most of the produced microsystems are made in silicon with different thinfilm processes. Technological procedures are well-engineered and a lot of research work is done to improve this kind of making microsystems. What could be the reason for using LTCC in microsystems to an increasing degree?

Possible structural sizes of microstructures made in silicon are smaller than 1 micron. But such a small size is not necessary for all microsystems. For a number of microsystems structural sizes .50 micron are needed. Bigger structure size and the consequential increasing total size of structures cause higher production costs. The production of "large" microsystems in silicon technology is only efficient up to a certain point. [9-11]

Microstructures in silicon are made with etching and deposition technologies. Therefore it is difficult to make structures with sufficient thickness for mechanical load capacity. The size and ruggedness of LTCC-based microstructures allows a simplified and cheaper handling and packaging. One packaging level is saved in best case [1].

### B. Changes in LTCC Technology

Using LTCC-Technology for microsystems requires some changes in technology. The changes affect both parameters and sequence of technological steps. The sequence of steps in the standard LTCC-Process is normally fixed. Making mechanical microstructures in LTCC requires a more complex procedure (Figure 3). The sequence of process steps depends on the manufactured structure and contains some loops. Table I shows some possibilities for mechanical processing. Because of the much better workability of LTCC in "green" state, mechanical processing after sintering should be avoided. The parameters of single steps also depend on the particular structures. Especially the lamination steps require special consideration because lamination using conventional parameters would destroy all concave structures inside the substrate and deform thin structures. Different from standard LTCC-Process more than one lamination step is needed in most cases. It is necessary to take the right choice between isostatic and uniaxial lamination. The lamination pressure can differ from the standard pressure very strongly. Lamination temperature and time are also to be regarded as variable values. The increase of the number of lamination steps causes a deterioration of the elastic properties of LTCC-Material. [12-14]

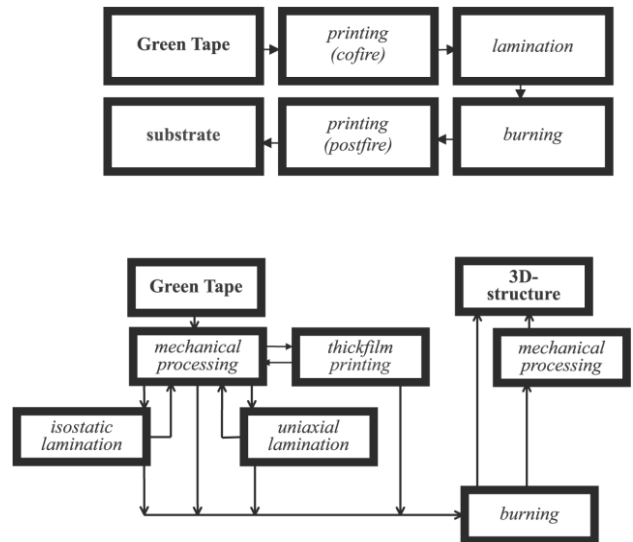

**Figure 3:** Processes for standard and advanced fabrication

**Table 1:** Mechanical processing

| Before burning | After burning |
|---|---|
| Laser cutting | Laser cutting |
| Punching | Sawing |
| Cutting | Drilling |
|  | Grinding |

### IV. DEVICES WITH LTCC

#### A. Chemical sensors

An electrochemical sensor was developed for application in Hot-Layer-Electrochemistry, a chemical experimental strategy which should permit observations of the electrochemical behavior of chemical compounds in supporting electrolytes at higher temperatures whilst minimizing the problems of bulk deterioration by fast heating [2]. The sensor uses direct heated electrodes. LTCC was selected as base material for several reasons. The ceramic has good resistance to several chemical solvents and permits the making of tight through holes for electrical connections. Hot-Layer- Electrochemistry requires the heating of the electrodes above the boiling point of the used solvent. Therefore a heat resistant material is needed. Another advantage is the very smooth surface which enables an easy chemical cleaning. Figure 4 shows the layout of the sensor. [15]

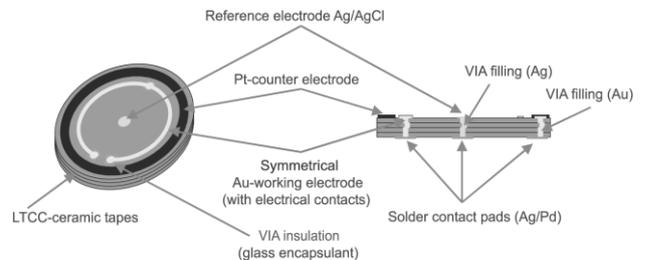

**Figure 4:** Direct heated chemical sensor.

Another sensor application is an indirect heated sensor [3]. It has two separate circuits for heating and for measurement. The described sensor has a thickness of only 200mm and consists of two layers of LTCC. The sensitive structure is accomplished by a printed interdigital structure and a sensitive layer made in thin film technology. [16]The distance between heater and sensor electrodes is as small as only 100mm and enables a fast dynamical behaviour of the sensor in association with a good thermal conductivity of the substrate material. Depending on the sensitive material that sensor can be used in temperature range up to 500C. Figure 5 shows a cross-section of an indirect heated sensor.

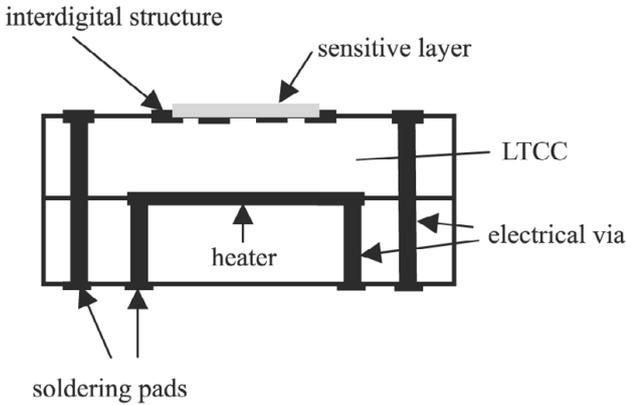

**Figure 5:** Indirect heated sensor

In future developments a number of similar sensors can be assembled with fluidic channels and other chemical and electrical components to more complex microsystems, which are comparable to "Lab on Chip"-designs in silicon technology.

### B. Fluidic channels and water cooling

LTCC offers the possibility to integrate channels into the substrate. The main problem is making such kind of structures is to find the best lamination parameters. The optimal size of lamination pressure depends on substrate thickness and channel width. [17-18] Figure 6 shows a cross section of an integrated channel. It is possible to accomplish a wide range of channel profiles. A possible application for that kind of channels is an integrated cooling system. It consists of microchannel patterns near the heat source. Figure 6 shows an ultrasonic microscope picture of a water cooled test structure. The brighter, meandered structure represents the fluidic channel. A printed resistor above this channel in the centre of the substrate works as a heater. Figure 7 shows the effect of an active cooling system on substrate temperature. Compared with passive cooling the temperature will be reduced very strongly [4].

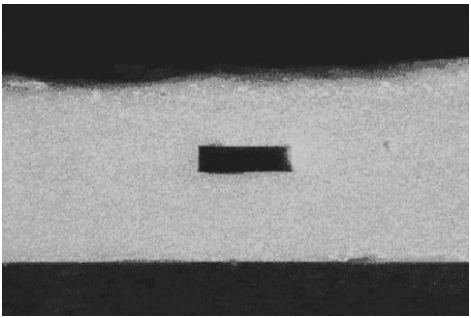

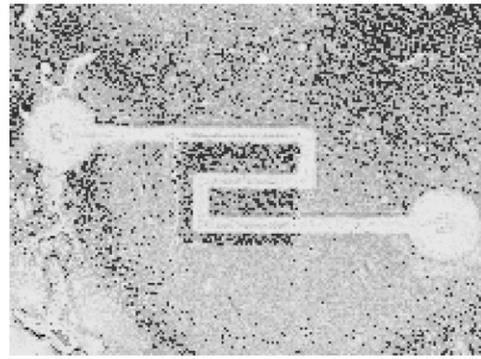

**Figure 6:** Microchannel in LTCC – cross section and top view

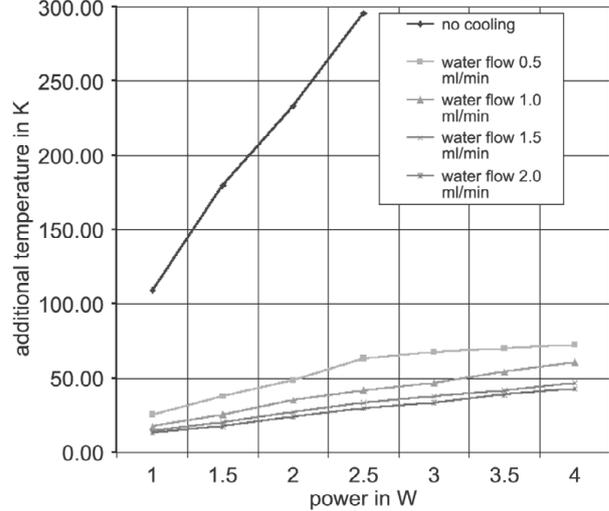

**Figure 7:** Effect of passive cooling (additional temperature = resistor temperature - room temperature)

### C. Cantilever Structures

LTCC-Technology offers a possibility to make ceramic cantilever structures or cantilever arrays by using only one or two layers of tape. These structures are to be found in some kind of mechanical sensors, for instance, to detect vibrations or accelerations (Figures 8-10). Compared with similar structures made in silicon, they are 100–1000 times bigger [19-22]. The outcome of this is other applications or other measurement ranges. Both, one-sided or two-sided fixed support is possible (Figure 10). The bigger size and stability of ceramic cantilever structures enables the use of bigger active masses. The conversion from mechanical to electrical values can be achieved by piezoelectric effect or by using the gauge behavior of printed resistors.

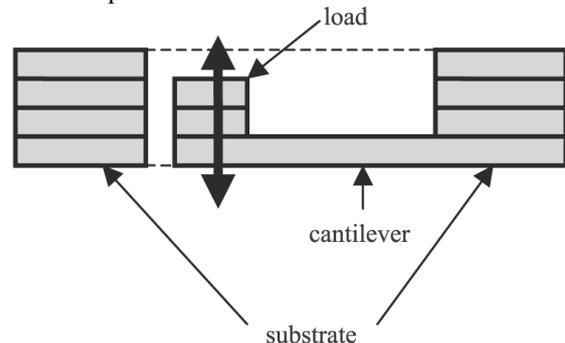

**Figure 8:** One-sided fixed cantilever (sketch)

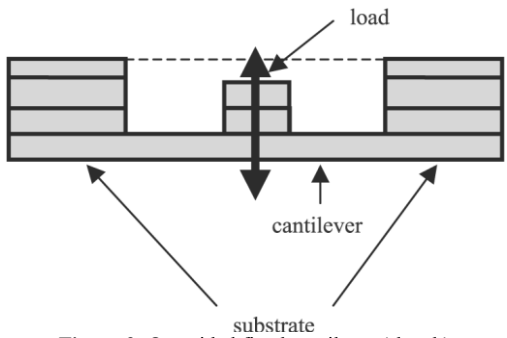

**Figure 9**: One-sided fixed cantilever (sketch)

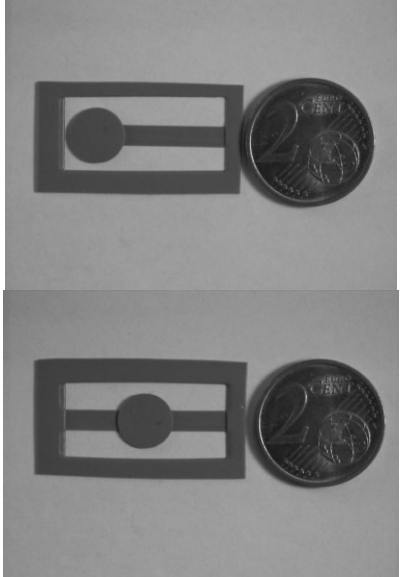

**Figure 10**: One-sided and two-sided fixed cantilever (photo)

### D. Micropump

Thin layers of LTCC represent flexible membranes. A passive system (sensor) containing such a structure can be used as pressure sensor, as force sensor or an active system (actuator) as pump. Figure 11 shows a cross section of a micropump made in LTCC-Technology. The picture shows a part of the membrane and a fluidic channel. The micropump is driven by a printed piezoelectric layer [5]. The thickness of the membrane is 130mm. The flow direction is determined by passive valves. The actual flow rate of the test examples averages approximately 0.7 ml/min. The ceramic parts of the micropump are structured completely before sintering, no additional bonding is needed (Figure 12 and 13) [23]. For this reason the manufacturing of this pump requires high demands on the lamination process [24]. In some parts two cavities are arranged one upon the other and passive valves and fluidic channels are also included. Therefore five different lamination steps are needed. Both isostatic and uniaxial lamination are used. The piezoelectric layer and the appropriate electrodes are made in a post fire process. The next development steps include testing of different drives and changing fluidic design in a way, that the pump works self-priming [25-27].

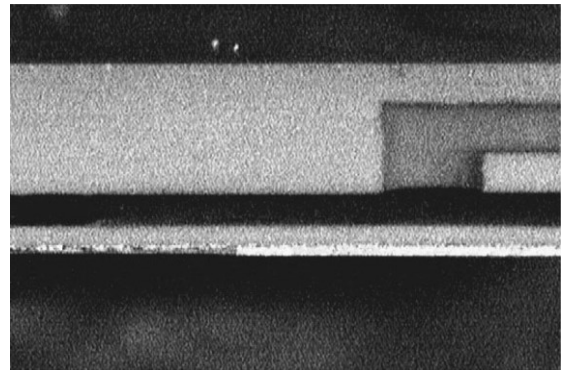

**Figure 11:** Cross section of LTCC-micropump

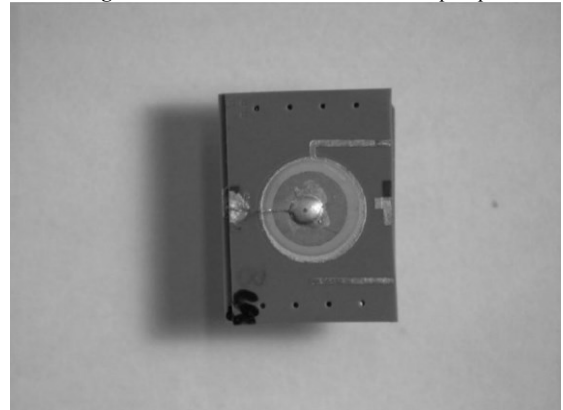

**Figure 12**: LTCC-micropump: piezoelectric drive

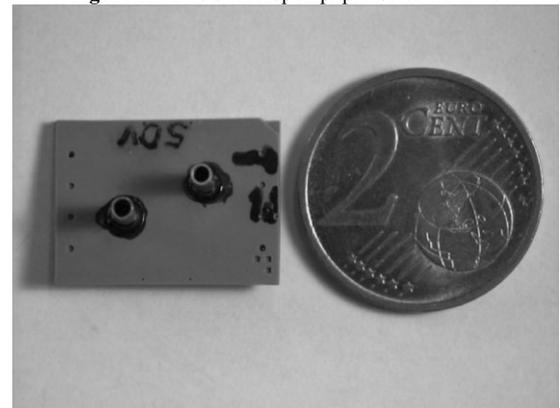

**Figure 13:** LTCC-micropump: fluidic connections

### V. CONCLUSION

The previously mentioned enhancements / improvements & cost drivers combined with the need to continually cut spending for non-recuning engineering (NRE) and a much more competitive supplier base worldwide, has forced a majority of the manufacturers, suppliers & users of electronic systems to seek ways to better utilize current as well as evolving technologies. Just to recap, they are higher performance with increased added value, higher compute power, faster compute operations, increased reliability, demands to reduced overall system costs, increased system usage world wide, continuing need to reduce overall product size and significantly reduced product life cycles This is the case with LTCC as a substrate for a multitude of interconnect & packaging system needs.

LTCC can be an interesting alternative as base material for microsystems. Perhaps it can replace other materials in still existing microsystem applications but mainly it should be used in microsystems which are designed specifically to use the advantages of the LTCC material system.

Some examples for non-electric functions and applications were shown in this paper. One task for future work is to improve the devices, but the main focus is the integration in more complex structures. A main problem is to find technology parameters and procedures, which take into account the different demands of several electrical, sensory and mechanical components and avoid negative interactions between then.

ACKNOWLEDGMENT

The authors would like to gratefully acknowledge the Microsystems Lab at the BMS College of Engineering for the support for their on going research in this field.